\documentclass[sigconf]{acmart}
\usepackage{verbatim}
\usepackage{amsmath}
\usepackage{listings}
\usepackage{xcolor}
\usepackage{caption}
\usepackage{setspace}

\usepackage{subcaption}
\PassOptionsToPackage{hyphens}{url}
\usepackage{booktabs}
\usepackage{pdfpages}
\setcopyright{none}
\renewcommand\footnotetextcopyrightpermission[1]{} 
\usepackage[font=small,labelfont=bf]{caption}

\begin{document}






%

\title{High Performance Data Persistence in Non-Volatile Memory \\ for Resilient High Performance Computing}
\author{Yingchao Huang}
\affiliation{%
  \institution{University of California, Merced}
}
\email{yhuang46@ucmerced.edu}

\author{Kai Wu}
\affiliation{
	\institution{University of California, Merced}
}
\email{kwu42@ucmerced.edu}

\author{Dong Li}
\affiliation{
	\institution{University of California, Merced}
}
\email{dli35@ucmerced.edu}

\begin{abstract}
Resilience is a major design  goal  for HPC. Checkpoint is the most common method to enable resilient HPC.  Checkpoint periodically saves critical data objects to non-volatile storage to enable data persistence. However, using checkpoint, we face dilemmas between resilience, recomputation and checkpoint cost. The reason that accounts for the dilemmas is the cost of data copying inherent in checkpoint. In this paper we explore how to build resilient HPC with non-volatile memory (NVM) as main memory and address the dilemmas. We introduce a variety of optimization techniques that leverage high performance and non-volatility of NVM to enable high performance data persistence for data objects in applications. With NVM we avoid data copying; we optimize cache flushing needed to ensure consistency between caches and NVM. We demonstrate that using NVM is feasible to establish data persistence frequently with small overhead (4.4\% on average) to achieve highly resilient HPC and minimize recomputation.
\vspace{-15pt}
\end{abstract}

\maketitle

\section{Introduction}
\label{sec:intro}

Resilience  is  one  of  the  major  design  goals  for extreme-scale HPC systems.  Looking forward to future HPC with shrinking feature size of hardware and aggressive
power management techniques, mean time between failures (MTBF) in HPC
could be shortened because of more frequent soft and hard errors;
the application execution could be interrupted more frequently;
and the application result correctness could be corrupted more often.

To address the above resilience challenge, checkpoint (or more specifically, application-level checkpoint) is the most common method deployed in current production supercomputers.
Application level checkpoint periodically saves application critical data objects to non-volatile storage 
to enable data persistence.
Once a failure happens, the application can restart from the last valid state of the data objects without restarting from the beginning.
However, checkpoint faces two dilemmas. 
First, there is a dilemma between HPC resilience and checkpoint overhead.
On one hand, as MTBF may become shorter in the future, 
we have to increase checkpoint frequency to improve HPC fault tolerance.
On the other hand, the frequent checkpoint results in larger runtime overhead.
We call this dilemma as the resilience dilemma.
Second, there is a dilemma between recomputation cost
and checkpoint overhead. 
On one hand, we want to increase checkpoint frequency to 
minimize recomputation cost and reduce data loss. 
On the other hand, the frequent checkpoint results in larger runtime overhead. We call this dilemma as the recomputation dilemma.

The fundamental reason that accounts for the above two dilemmas 
is the cost of data copying inherent in the checkpoint mechanism.
The data copying operations can be expense, because
checkpoint data has to be stored in remote or local durable hard drive. 
Although the disk-less checkpoint reduces data copying overhead~\cite{tpds98:plank, Lu:2005:SDC:1145057, ppopp17:tang, isftc94:plank, ipdps09:bronevetsky}
by using main memory, this technique has to encode data across multiple nodes to create redundancy and only tolerates up to a certain number of node failures, because of the volatility of memory.
Other techniques, such as multi-level checkpoint~\cite{sc10:moody, sc09:dong, sc11:gomez} and incremental checkpoint~\cite{isftc94:plank, ics04:agarwal, icpads10:wang, ipdps09:bronevetsky} partially remove expensive data copying off the critical path of application execution,
but a checkpoint with a large data size can still cause large runtime overhead.

The emergence of non-volatile memories (NVM), such as phase change memory (PCM) and RRAM, is poised to revolutionize memory systems~\cite{sc10:Caulfield, cse15:vetter}.
The performance of NVM is much better than hard drive, and even close to or match that of DRAM~\cite{NVMDB, eurosys16:dulloor}. 
Furthermore, NVM has better scalability than DRAM while remain non-volatility. These features make it possible to merge the traditional two layers of memory hierarchy (i.e, memory plus back-end storage) into one layer (i.e, memory without back-end storage)~\cite{imw13:mutlu}.
Given NVM as main memory and its non-volatility nature,
is it possible to change or even remove checkpoint to enable data persistence frequently, thus fundamentally addressing the above two dilemmas in future HPC systems?
How can NVM be used to address the resilience challenge for HPC?

This paper aims to answer the above questions, and explores
how to build resilient HPC with emerging NVM as main memory.
We introduce a variety of optimization techniques
to leverage high performance and non-volatility of
NVM to establish data persistence for application critical data objects frequently.

We start from a preliminary design that uses NVM as either main memory or storage to implement checkpoint. We expect that the superior performance of NVM would allow us to achieve frequent checkpoint with small runtime overhead and hence address the two dilemmas. 
To improve checkpoint performance, we introduce 
a couple of optimizations, including parallelization of cache flushing and using 
SIMD-based, non-temporal load/store instructions (e.g.,
MOVDQU) to bypass CPU caches and minimize data movement between caches
and memory.
However, we reveal that even based on an optimistic assumption on NVM performance, NVM-based checkpoint can still lead to large runtime overhead (up to 46\%), because of data copying in checkpoint. 

We further study how to leverage non-volatility of NVM to create a copy
of the data objects. We aim to replace traditional data copying in checkpoint, which is the fundamental reason that accounts for expensive checkpoint. 
We introduce a technique, named in-place versioning.
This technique hides programmers from application and algorithm details, and leverages application-inherent memory write operations to create a new version of the data objects in NVM without extra data copying. We derive a set of rules to enable automatic transformation of programs to achieve in-place versioning. 

To ensure proper recovery based on the new version of the data objects, 
we must guarantee that the data of the new version is consistent between caches and NVM.
Hence, we must flush data blocks of the new version out of caches, after the new version is created by the in-place versioning technique.
Such cache flushing operations can be expensive, because there is
no mechanism that allows us to track which data blocks of the new version are in caches and whether data blocks in caches are clean. As a result, we must flush all data blocks of the new version as if all data blocks are in caches,
which brings large performance loss.

To minimize the cache flushing cost, we propose to use a privileged instruction and make it accessible to the application to flush the entire cache hierarchy, instead of flushing all data blocks of the new version. For a large data object, flushing the entire cache hierarchy are often much cheaper.
Furthermore, we introduce an asynchronous and proactive cache flushing mechanism to 
remove cache flushing cost off the critical path of application execution
while enabling data consistency in NVM.

In general, the in-place versioning plus the optimized cache flushing allow us to establish data persistence with consistence for application critical data objects in NVM.  
The establishment of data persistence can happen much more frequently 
than the traditional checkpoint mechanism, with high performance.
With the evaluation of six representative HPC benchmarks and one production HPC application (Nek5000), we show that the runtime overhead is \%4.4 on average (up to 9\%) when the establishment of data persistence frequently happens at every iteration of the main computation loop. 
Such frequent and high performance data persistence allows us to
minimize recomputation cost and tolerate high error rate
in future HPC.

Our major contributions are summarized as follows.
\vspace{-1pt}
\begin{itemize}
  
  \item We explore how to use NVM to enable resilient HPC. We demonstrate that using NVM (either as main memory or storage) to implement frequent checkpoint based on data copying to address the two dilemmas may not be feasible, because of large data copying overhead, even though NVM is expected to have superior performance.
  
  \item We explore how to enable data persistence with consistency in NVM with minimized runtime overhead. Without data copying and with the optimization of cache flushing, using NVM has potential to address the resilience and recomputation dilemmas rooted in the traditional checkpoint.
\end{itemize}

\section{Background}
\label{sec:bg}
In this paper, we focus on HPC applications.
Those applications are typically characterized with iterative structures.
In particular, there is usually a main computation loop in an HPC application. 
With the traditional checkpoint mechanism, at every $n$ iterations of the loop ($n$ is much larger than 1), the application saves critical data objects of the application into non-volatile storage. 
In the rest of the paper, we name those critical data objects as \textit{target data objects}. 
Checkpoint usually happens near the end of an iteration.
We call the execution point where checkpoint happens as \textit{persistence establishment point}. 

We also distinguish \textit{cache line} and \textit{cache block} in this paper.
The cache line describes a location in the cache, and the cache block
refers to the data that go into a cache-line. 
We review NVM background in this section.


\subsection{Non-Volatile Memory Usage Model}
There are at least two existing usage models to integrate the emerging NVM into 
HPC systems. 
In the first model, NVM is built as NVDIMM modules and installed 
into DDR slots. NVM is physically attached to the high-speed memory bus and managed by a memory controller~\cite{micro16:chen}. 
In the second model, NVM connects to the host
by an I/O controller and I/O bus (e.g., PCI-E)~\cite{msst14:chen}.  

From the perspective of software, OS can regard NVM as regular memory (the first model), similar to DRAM, and NVM provides the capability of being byte addressable to OS and applications. Also, NVM is accessed through {\fontfamily{qcr}\selectfont
load} and {\fontfamily{qcr}\selectfont store} instructions.
Alternatively, NVM can be exposed as a block device in OS~\cite{usenix13:rudoff}. 
NVM is accessed via a read/write block I/O interface. 
A file system can be built on top of NVM
to provide the convenience of naming schemes and data protection~\cite{usenix13:rudoff}.
\subsection{Data Consistence in NVM}
To build a consistent state for target data objects in NVM (as main memory) and ensure proper recovery, 
the target data objects in NVM must be updated with the most
recent data in caches at the persistence establishment point.
However, the prevalence of 
volatile caches introduces randomness into write operations in NVM.
When the data is written from caches to NVM is subject to
the cache management policy by hardware and OS.

There are ``interfaces'' that enable explicit data flushing 
from caches to NVM. Those interfaces are presented as processor instructions
or system calls. 
Using those interfaces, it is possible to enforce data consistence at
the persistence establishment point.
We discuss the common cache flushing instructions as follows. 

\vspace{-5pt}
\begin{itemize}
\item {\fontfamily{qcr}\selectfont clflush} instruction: 
This is the most common cache flushing instruction. 
Given a cache block, this instruction invalidates it from all levels of the processor cache hierarchy. If the cache line at any level of the cache hierarchy is dirty, the cache line is written to memory before invalidation. 
{\fontfamily{qcr}\selectfont clflush} is a blocking instruction, meaning that the instruction waits until the data flushing is done~\cite{nvmsummit16:rudoff}. 

\item {\fontfamily{qcr}\selectfont WBINVD} instruction: this is a privileged instruction used by OS to flush and invalidate the entire cache hierarchy.

\end{itemize}

To enable data consistence based on {\fontfamily{qcr}\selectfont clflush} and other cache block-based cache flushing instructions (particularly {\fontfamily{qcr}\selectfont CLWB} and {\fontfamily{qcr}\selectfont clflush\_opt}, which will be discussed next),
we may have a performance problem 
for a data object with a large data size.  
Because we do not have a mechanism to track which cache line is dirty and whether a specific cache block is in caches, we have to flush all cache blocks of target data objects, as if all cache blocks are in caches. Figure~\ref{fig:cache_flushing} shows how we flush cache blocks based on cache block-based cache flushing. 

\definecolor{codegreen}{rgb}{0,0.6,0}
\definecolor{codegray}{rgb}{0.5,0.5,0.5}
\definecolor{codepurple}{rgb}{0.58,0,0.82}
\definecolor{backcolour}{rgb}{0.95,0.95,0.92}
\lstdefinestyle{style1}{
    commentstyle=\color{codegreen},
    keywordstyle=\color{magenta},
    numberstyle=\tiny\color{codegray},
    stringstyle=\color{codepurple},
    basicstyle=\footnotesize,
    numbers=left,                    
    numbersep=5pt, 
	escapeinside={(*@}{@*)},
}
\lstset{style=style1}

\begin{figure}[!htb]
\centering
\begin{lstlisting}[language=c]

 /*Loop through cache-line-size aligned chunks
covering the given range of the target data object*/
cache_block_flush(const void *addr, size_t len)
{
  unsigned __int64 ptr;
        
  for (ptr = (unsigned __int64)addr & ~(FLUSH_ALIGN - 1);
       ptr < (unsigned __int64)addr + len; 
       ptr += FLUSH_ALIGN)
       flush((char *)ptr);  /*clflush/clflush_opt/clwb*/
}
\end{lstlisting}
\vspace{-10pt}
\caption{Using cache block-based cache flushing instructions to flush cache blocks of the target data object.}
\label{fig:cache_flushing}
\vspace{-10pt}
\end{figure}

Flushing clean cache blocks in caches and flushing cache blocks not in caches have performance cost at the same order as flushing dirty cache blocks.
Table~\ref{tab:diff_clflush_perf} shows the performance of flushing cache blocks in different status in caches. The performance is measured in a platform with two eight-core Intel Xeon E5-2630 v3 processors (2.4 GHz, 20MB L3, 256KB L2, and 32KB L1) attached to 32GB DDR4. Based on the results, we conclude that flushing all cache blocks of a data object is roughly proportional to the data object size.

\begin{table}
\centering
\caption{Performance of flushing cache blocks in different status in caches using {\fontfamily{qcr}\selectfont clflush}.}
\vspace{-10pt}
\scriptsize 
\begin{tabular}{| p{1.7cm} | p{1.7cm} | p{1.7cm} | p{1.8cm} |}
       \hline
                & \textbf{Flush dirty cache blocks in caches} & \textbf{Flush clean cache blocks in caches} & \textbf{Flush cache blocks not in caches} \\ \hline \hline
    Cycles per cache block & 228 & 254 & 350 \\ \hline
\end{tabular}
\label{tab:diff_clflush_perf}
\vspace{-10pt}
\end{table}

To support NVM, there are two very new instructions, {\fontfamily{qcr}\selectfont clflush\_\\opt} and {\fontfamily{qcr}\selectfont CLWB}.
{\fontfamily{qcr}\selectfont clflush\_opt} maximizes the concurrency of multiple {\fontfamily{qcr}\selectfont clflush} within individual threads. {\fontfamily{qcr}\selectfont CLWB} instruction maximizes the concurrency of
multiple cache line flushing without cache line invalidation (i.e., leaving data in the cache after cache line flushing).
{\fontfamily{qcr}\selectfont clflush\_opt} is only available in the most recent Intel SkyLake microarchitecture. Based on our knowledge,
there is no hardware available in the market that supports {\fontfamily{qcr}\selectfont CLWB}.
We cannot evaluate them in this paper. However,
using these two instructions should lead to better performance 
with our method proposed in this paper. More importantly, these two instructions use cache block-based cache flushing, hence they have the same problem as discussed above for large target data objects. Our proposed method can help them improve performance.



\section{Preliminary System Designs}
\label{sec:prelim_design}
The performance of NVM is much better than that of traditional hard drive, and even close to or match that of DRAM. 
Given such performance characteristics of NVM, it is promising to enable 
frequent checkpoint with a small overhead.
Frequent checkpoint will enable better HPC resilience and minimize recomputation, hence addressing the two dilemmas for future HPC.

Our preliminary designs aim to improve the existing checkpoint mechanism and optimize its performance on NVM. We want to answer a fundamental question: can the NVM-based checkpoint (with optimization) happen frequently, such that we address the two dilemmas rooted in the current checkpoint mechanism?
 
\subsection{Preliminary Design 1: NVM-based, Frequent Checkpoint}
In our first design, we employ an NVM-based checkpoint, and the checkpoint
happens at each iteration of the main loop, which is much more frequent
than the traditional checkpoint.
Also, the NVM-based checkpoint happens locally. This means that no matter what usage model NVM is used (either as main memory or as a local block device), the checkpoint is stored locally in NVM. By removing networking overhead, this local NVM-based checkpoint represents the best performance we can get out of NVM.
In fact, from the~\textit{architecture point} of view, such local NVM-based system has been shown to be possible for HPC~\cite{ipdps13:kannan, ics15:gao, sc13:jung}.

We compare two cases of hard drive-based, frequent checkpoint
with two cases of NVM-based, frequent checkpoint. 
For hard drive-based, frequent checkpoint, 
the hard drive is resident either locally (annotated as ``hard drive based chkp (local)'') or in a remote storage node (annotated as ``hard drive based chkp (remote)''). 
For NVM-based, frequent checkpoint, NVM is used 
as either main memory (annotated as ``NVM based chkp (mem)'')
or a local block device (annotated as ``NVM based chkp (block)'').
If NVM is used as main memory, checkpointing is the same as
making a data copy in memory plus necessary cache flushing.
To emulate NVM as a block device, we use a ramdisk with a file system
(tmpfs). Hence, such emulation includes the overhead of file system and system calls, but does not emulate internal overhead of I/O controllers, such as interface command decoding and ECC. 

We run six NAS parallel benchmarks and one production code (Nek5000). The details for those applications are summarized in Table 2. 
In our study, NVM has either the same performance characteristics (bandwidth and latency) as DRAM, which is a rather optimistic assumption on NVM performance, or inferior performance than DRAM, which is a more practical assumption. 

\textbf{(1) NVM has the same performance as DRAM.} 
We emulate NVM with local DRAM, similar to~\cite{asplos15:zhang} and assume that NVM has the same latency and bandwidth as DRAM, 
After data copying in checkpoint, we flush cache blocks of the new data copy out of caches to build a consistent state in NVM, using {\fontfamily{qcr}\selectfont clflush}.

\begin{figure}
\centering
\includegraphics[width=0.48\textwidth, height=0.15\textheight]{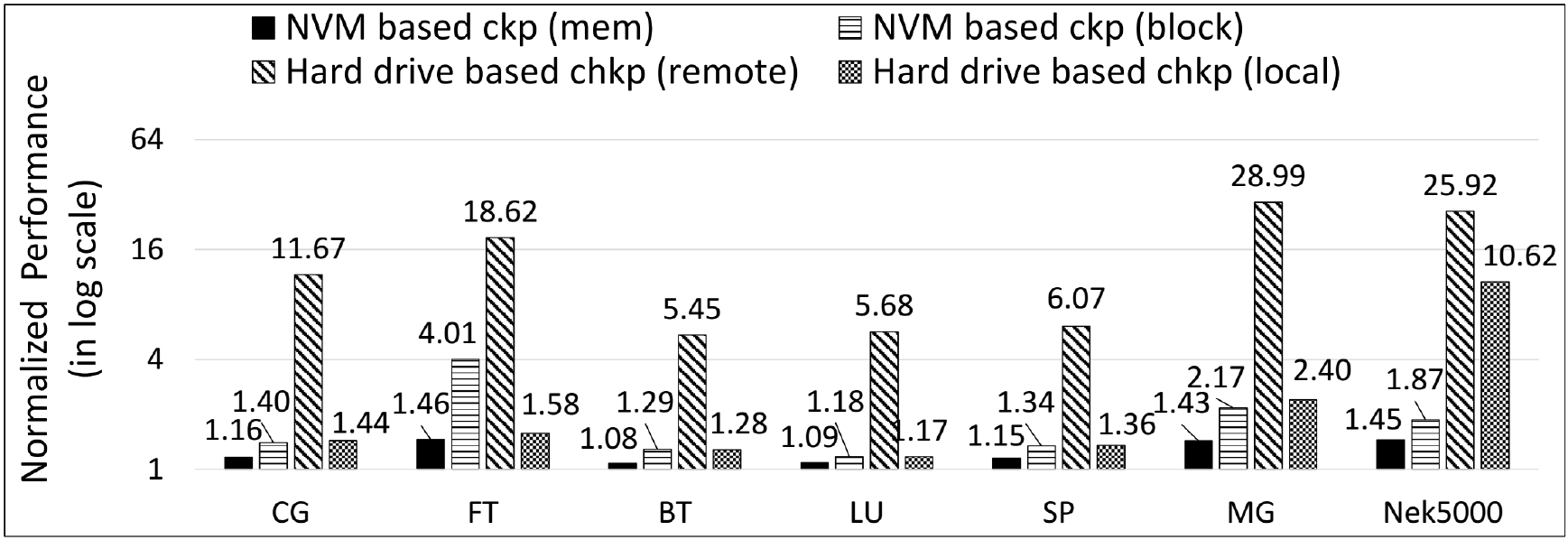}
\vspace{-20pt}
\caption{Preliminary design 1 with NVM-based, frequent checkpoint and hard drive-based, frequent checkpoint on Edison. NVM has the same performance characteristics as DRAM. Performance (execution time) is normalized by that of native execution on DRAM without checkpoint.}
\label{fig:prelim_design1}
\vspace{-15pt}
\end{figure}

Figure~\ref{fig:prelim_design1} shows the results on a production supercomputer, Edison at Lawrence Berkeley National Lab. 
For NPB benchmarks, we use CLASS D as input; for Nek5000, we use the eddy problem as input ($256 \times 256$). 
We use 4 nodes with 16 MPI tasks per node.
Performance (execution time) in the figure is normalized by that of the native execution without checkpoint. 
The figure reveals that with frequent checkpoint, hard drive based checkpoint (local) has 283\% overhead on average (up to 1062\%), which is unacceptable.
NVM-based checkpoint has much better performance.
For some benchmarks (e.g., BT and LU), the overhead of NVM-based checkpoint (NVM as main memory) is smaller than 10\%.
But there is still high overhead for some benchmarks (more than 40\% for MG, FT, and Nek5000).
Also, NVM-based checkpoint (main memory) shows better performance (26\% performance loss on average and up to 46\%) than NVM-based checkpoint (block device) (89\% performance loss on average and up to 401\%).

\begin{figure}
\centering
\includegraphics[width=0.48\textwidth, height=0.13\textheight]{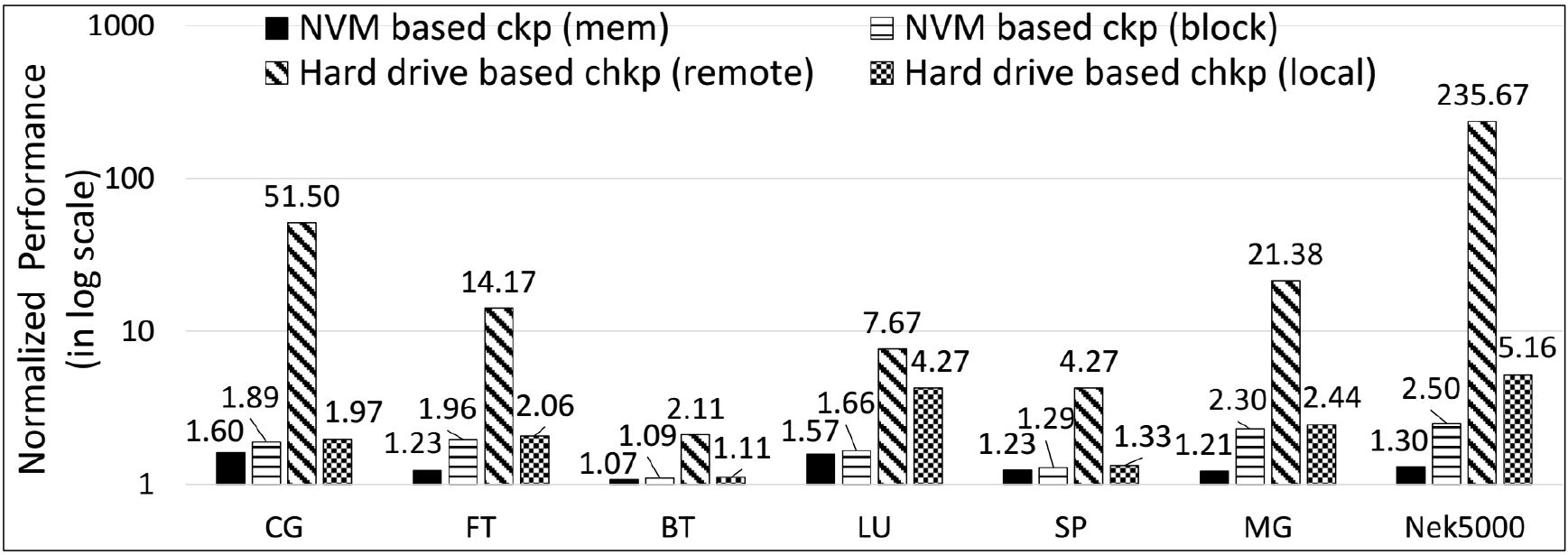}
\vspace{-20pt}
\caption{Preliminary design 1 with NVM-based, frequent checkpoint and hard drive-based frequent checkpoint on a local cluster. Performance (execution time) is normalized to that of the native execution without checkpoint on the heterogeneous NVM/DRAM system. NVM has 1/8 bandwidth of DRAM.}
\label{fig:prelim_design1_low_nvm}
\vspace{-10pt}
\end{figure}

\begin{figure}
\centering
\includegraphics[width=0.48\textwidth, height=0.13\textheight]{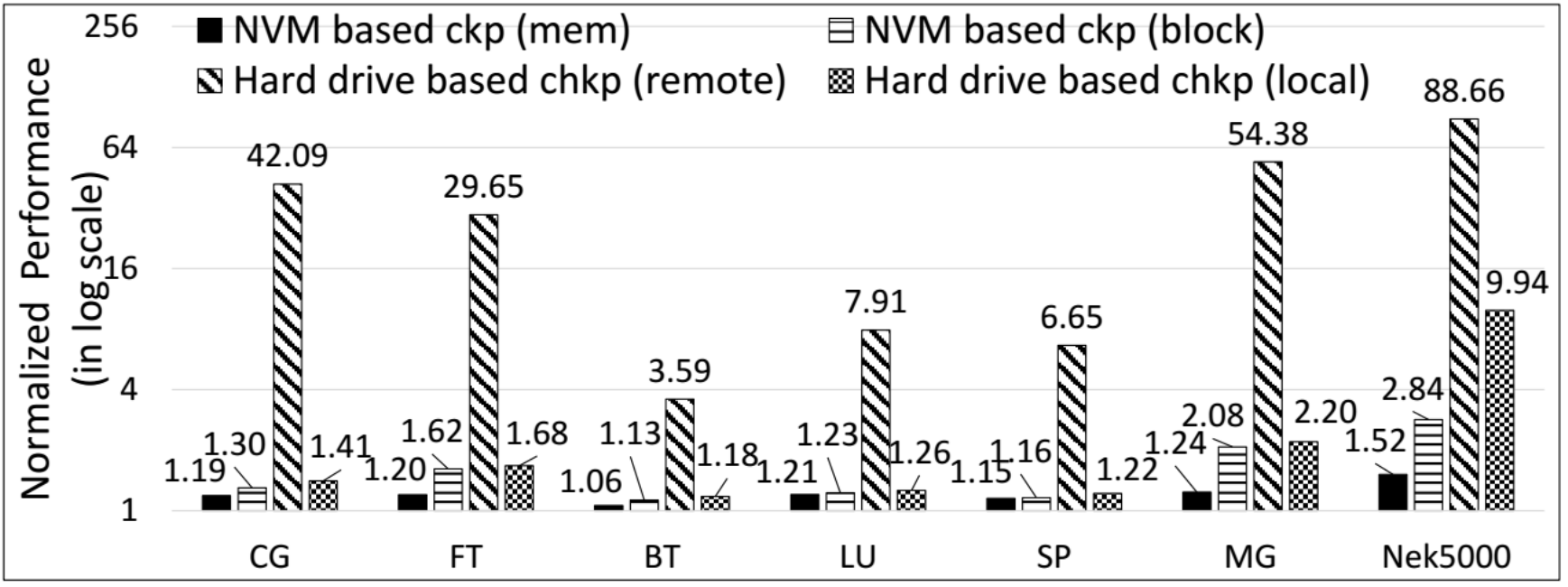}
\vspace{-20pt}
\caption{Preliminary design 1 with NVM-based, frequent checkpoint and hard drive-based frequent checkpoint on a local cluster. Performance is normalized to that of the native execution without checkpoint on the heterogeneous NVM/DRAM system. NVM has 1/32 bandwidth of DRAM.}
\label{fig:prelim_design1_low_nvm2}
\vspace{-10pt}
\end{figure}

\textbf{(2) NVM has worse performance than DRAM.}
Since the NVM techniques have a range of performance characteristics,
we change NVM performance to make our evaluation more practical, and re-do the above tests in (1). 
Since checkpoint performance is sensitive to memory bandwidth, we change NVM bandwidth based on Quartz (a DRAM-based, lightweight performance emulator for NVM~\cite{middleware15:volos}) for our study. Because using Quartz requires loading a kernel driver, which needs privileged accesses to the system, we run Quartz on a local cluster (see Section~\ref{sec:eval} for more details on the cluster). 
We choose 1/8 and 1/32 DRAM bandwidth as NVM bandwidth based on~\cite{eurosys16:dulloor, NVMDB}.
We use CLASS C as input for NPB and the eddy problem ($256\times256$) as input for Nek5000; we use 4 nodes with 4 MPI tasks per node.
Figures~\ref{fig:prelim_design1_low_nvm} and~\ref{fig:prelim_design1_low_nvm2} show the results. 

Note that given a lower NVM bandwidth, the application performance on a NVM-only system is worse than a DRAM-only system.
To bridge the performance gap between NVM and DRAM, the existing work 
introduces a small DRAM cache~\cite{pm_iccd14,Ramos:ics11, eurosys16:dulloor} to place recent write-intensive data into NVM and build
a heterogeneous NVM/DRAM system.
To study the impact of such small DRAM cache on checkpoint performance,
we allocate a small DRAM space to implement a heterogeneous NVM/DRAM system based on Quartz.
The existing work chooses the DRAM cache size between 32MB and 1GB~\cite{pm_iccd14,Ramos:ics11, eurosys16:dulloor, gpu_pcm_pact13, ieee_micro11:jiang, dac09_pdram, hpdc16:wu}.
We choose a medium DRAM size in our test, which is 256MB.

With the DRAM cache, the overhead of NVM-based checkpoint (NVM as main memory) must include flushing cache blocks of the target data objects from
this DRAM space to NVM, besides the overhead of memory copying in NVM and CPU cache flushing. 
Which data objects are in the DRAM cache at the persistence establishment point depends on the DRAM cache management strategy.
We implement a recent software-based approach~\cite{eurosys16:dulloor} to manage the DRAM cache.
Furthermore, because of the software-based approach, 
we know which target data objects (or data blocks of the target data objects) are in the DRAM cache. Hence, we do not need to flush all cache blocks of target data objects for DRAM cache flushing. Also, we do not invalidate data in the DRAM cache after DRAM cache flushing to optimize performance of DRAM cache flushing.

Figures~\ref{fig:prelim_design1_low_nvm} and~\ref{fig:prelim_design1_low_nvm2}
show the results. 
Similar to Figure~\ref{fig:prelim_design1}, the two figures show that NVM-based,
frequent checkpoint (NVM as main memory) can result in large performance loss (22\% on average and up to 52\% for NVM with 1/8 DRAM bandwidth, and 32\% on average and up to 60\% for NVM with 1/32 DRAM bandwidth).

\textbf{Conclusions.} Using NVM as main memory for checkpoint is promising, but still comes with large performance overhead for some benchmarks, even though we take an optimistic assumption on NVM performance. 

The performance loss of NVM-based checkpoint (NVM as main memory)
comes from data copying during checkpointing and cache flushing. 
To improve the performance of NVM-based checkpoint (NVM as main memory), we focus on improving the performance of cache flushing in the next section. We consider removing data copying in Section~\ref{sec:design}. 
In the rest of this paper, we focus on NVM with 1/8 DRAM bandwidth, which is a more practical assumption on NVM performance~\cite{NVMDB, eurosys16:dulloor}.

\subsection{Preliminary Design 2: Optimization of NVM-based Checkpoint}
\begin{figure}
\centering
\includegraphics[width=0.48\textwidth, height=0.13\textheight]{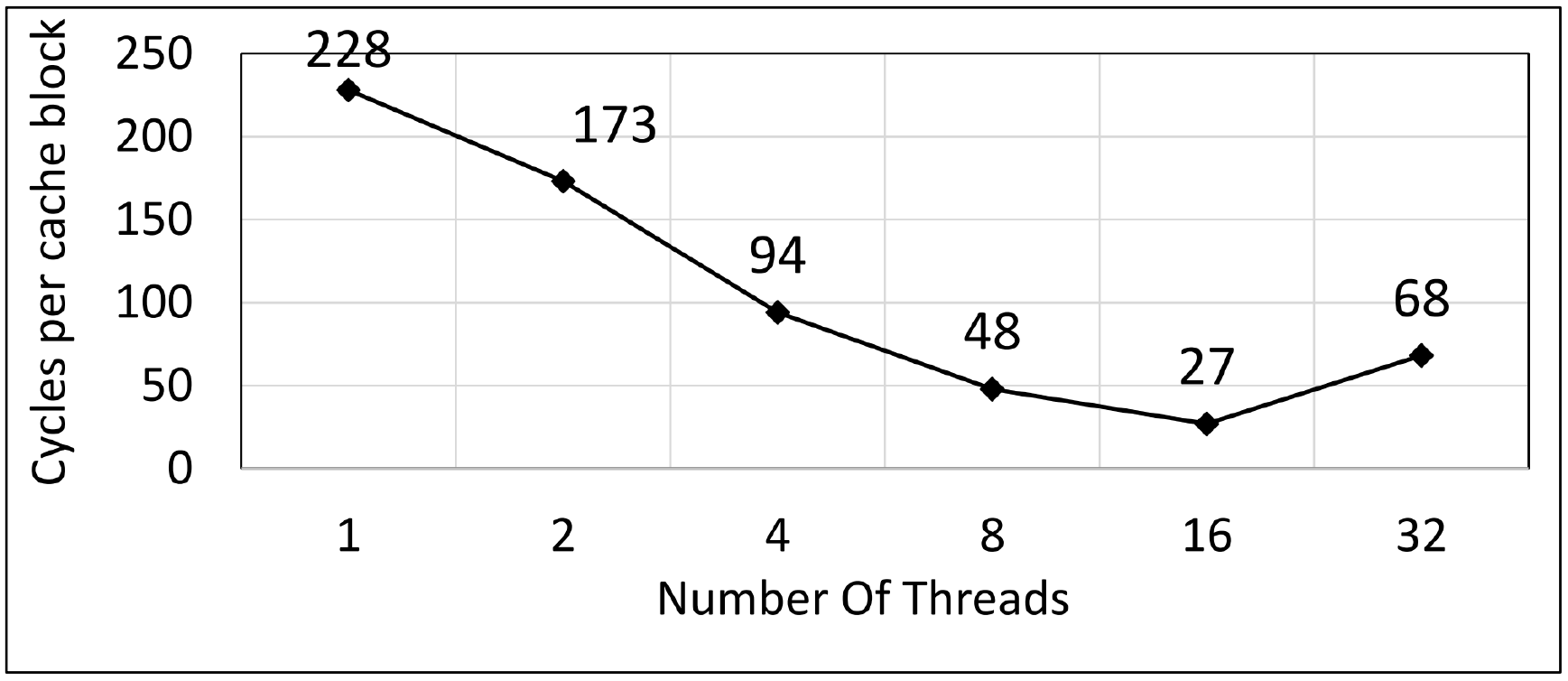}
\vspace{-20pt}
\caption{Performance of parallelizing {\fontfamily{qcr}\selectfont clflush} with multi-threading.}
\label{fig:clflush_perf}
\vspace{-10pt}
\end{figure}

To improve the performance of cache flushing, we explore 
the parallelization of {\fontfamily{qcr}\selectfont clflush} instructions by multi-threading.  
Although {\fontfamily{qcr}\selectfont clflush} is blocking, 
there is no guaranteed order for {\fontfamily{qcr}\selectfont clflush} instructions~\cite{clflush} across threads.
It is possible to use multiple threads for cache flushing, and
each of which flushes non-overlapped cache blocks.
To verify the above idea, we use OpenMP {\fontfamily{qcr}\selectfont parallel for} to parallel a \textit{for} loop for cache flushing with each iteration of the loop flushing a single cache block.
We change the number of threads and measure performance for flushing a 20MB data buffer with dirty cache blocks on an Intel Xeon E5-2630 v3 processor (20MB L3, 256KB L2, and 32KB L1)
attached to 32GB DDR4. The processor has 8 cores with 16 hardware threads.
Figure~\ref{fig:clflush_perf} shows the performance (average cycles per cache line).

Figure~\ref{fig:clflush_perf} shows that using multi-threading
does improve performance of {\fontfamily{qcr}\selectfont clflush}, but the performance is not scalable beyond certain number of threads.
In fact, as we increase the number of threads, they will compete for
those resources in cache controllers and read/write ports of main memory,
which limits the scalability of parallel {\fontfamily{qcr}\selectfont clflush}. Based on such observation, we use up to 16 threads to 
parallelize cache flushing, depending on the availability of idling cores
in a node.

To further improve performance of NVM-based checkpoint (NVM as main memory), 
we explore special instructions and use SIMD-based, non-temporal instructions (particularly MOVDQU and \\MOVNTDQ), which bypass caches
to make a data copy. Using those instructions removes the necessity of cache flushing, but those instructions are only available on
a processor with SSE support. 

\begin{figure}
\centering
\includegraphics[width=0.48\textwidth, height=0.13\textheight]{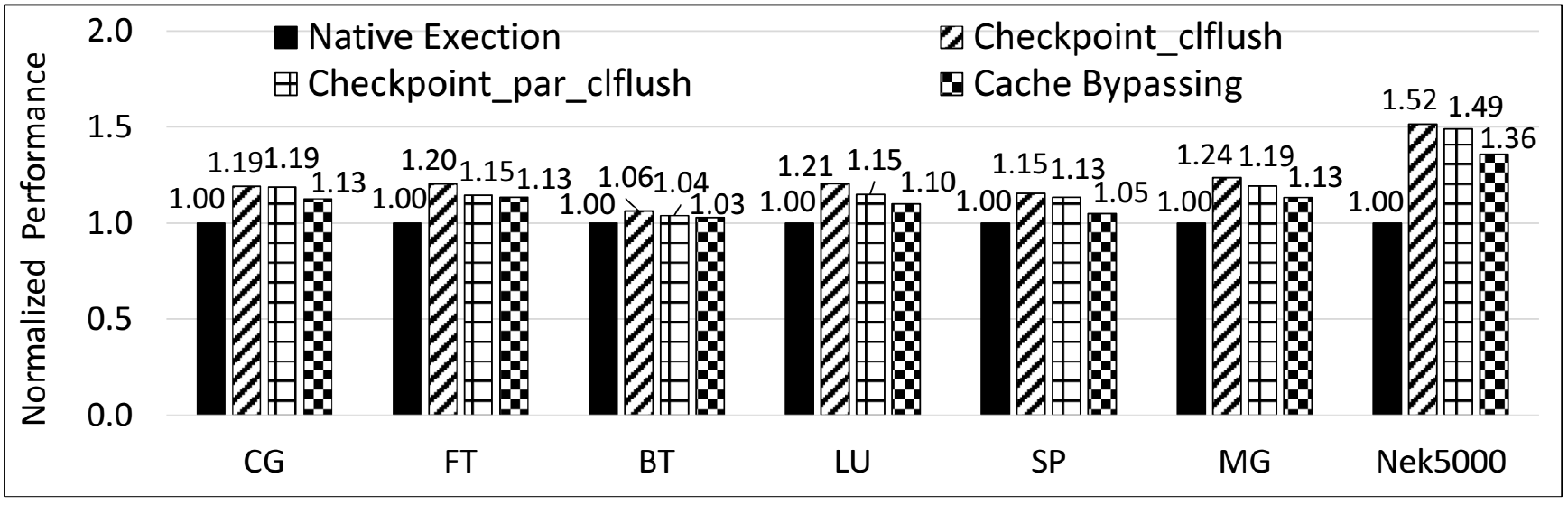}
\vspace{-20pt}
\caption{Performance (execution time) of NVM-based checkpoint with optimization (NVM is used as main memory). Performance is normalized by that of the native performance without checkpoint.}
\label{fig:prelim_design2}
\vspace{-20pt}
\end{figure}

Figure~\ref{fig:prelim_design2} shows the performance for the above two optimization techniques. Within the figure, the preliminary design 1 (i.e., NVM-based checkpoint with NVM as main memory),
the parallelized {\fontfamily{qcr}\selectfont clflush}, and non-temporal instructions are labeled as ``checkpoint\_clflush'', ``checkpoint\_par\_clflush'', and ``cache bypassing'' respectively. 
``Native execution'' is the one without checkpoint.

The figure shows that the parallelized {\fontfamily{qcr}\selectfont clflush} has up to 5\% performance improvement (for FT) over the preliminary design 1. Non-temporal instructions lead to the best performance in all cases.
Comparing with the preliminary design 1 (checkpoint\_clflush  in Figure~\ref{fig:prelim_design2}), non-temporal instructions result in 9.6\% performance improvement on average and up to 16\%. 
If a platform supports those instructions, they should be the preferred method for NVM-based checkpoint. 

\begin{figure*}
\centering
\includegraphics[width=1.0\textwidth, height=0.2\textheight]{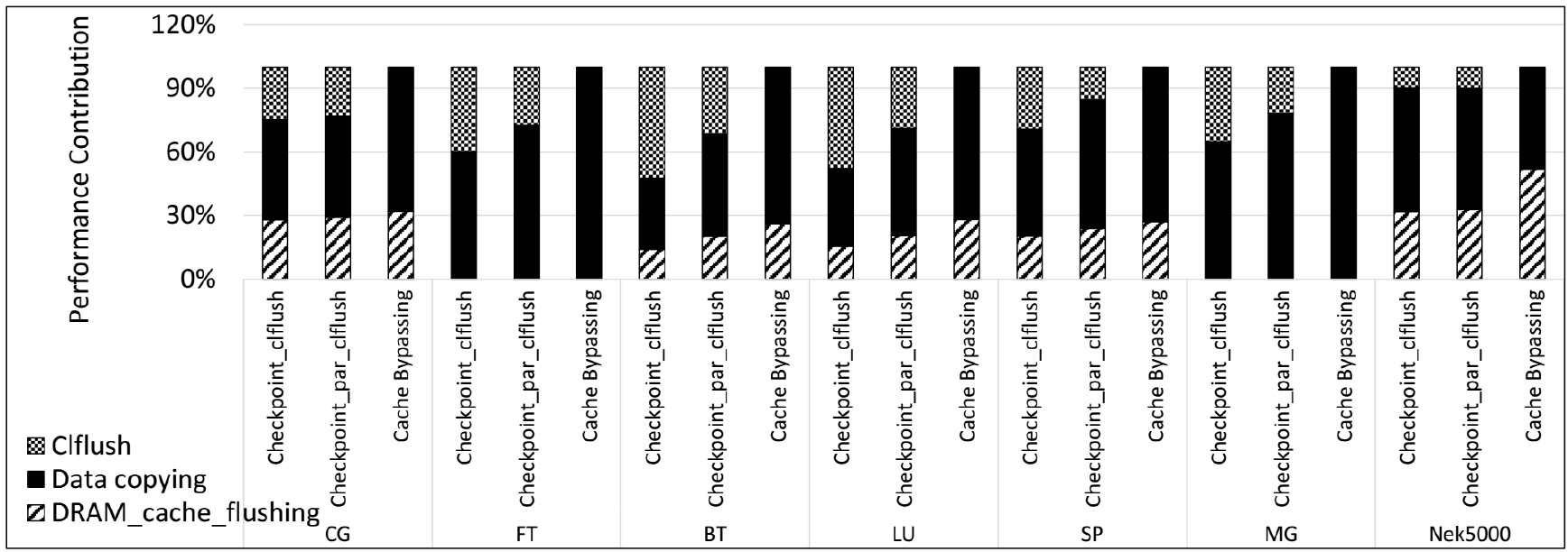}
\vspace{-20pt}
\caption{The breakdown of performance loss for NVM-based checkpoint after optimization (i.e., the preliminary design 2).}
\label{fig:perf_loss_breakdown}
\vspace{-10pt}
\end{figure*}

However, even if we use the above optimizations on CPU cache flushing,
we still see big performance loss on some benchmarks (e.g., 36\% for Nek5000 and 13\% for CG).
To investigate the reason, we break down the checkpoint time.
For ``checkpoint\_clflush'' (the preliminary design 1) and ``checkpoint\_par\_clflush'', the checkpoint time includes DRAM cache flushing, data copying, and CPU cache flushing; for ``cache bypassing'', the checkpoint time includes DRAM cache flushing and data copying. Figure~\ref{fig:perf_loss_breakdown} shows the results. 

The results reveal that data copying contributes the most
to the performance loss. Except BT and LU with the preliminary design 1, all other cases have more than 50\% performance loss come from data copying. 

\textbf{Conclusions.} To establish frequent data persistence in NVM with high performance and address the dilemmas in checkpoint, we must address the data copying overhead.

\section{High Performance Data Persistence}
\label{sec:design}
We introduce a technique, called ``in-place versioning'', to remove data copying. 
Because the in-place versioning has to come with cache flushing, 
we introduce an asynchronous and proactive
cache flushing 
to improve performance.

\subsection{In-Place Versioning}
\textbf{Basic Idea.}
The in-place versioning is based on the idea of the dual version~\cite{hpdc16:wu}.
Both the in-place versioning and the dual version aim to remove data copying by leveraging application-inherent memory write operations to create a new version of the target data objects. But the dual version heavily relies on numerical algorithm knowledge, and is only applicable to those algorithms with specific characteristics.
The implementation of the dual version for an algorithm requires the programmer to manually change the code based on algorithm knowledge.

The in-place versioning significantly improves the dual version.
The in-place versioning works for any numerical algorithm, and is algorithm-agnostic. We generalize a couple of rules to implement the in-place versioning.
Based on the rules, we can use compiler to automatically transform the application into a new one with the implementation of in-place versioning. The new application creates data copy at runtime without programmer intervention. In the following, we describe the basic idea of the dual version in an \textit{algorithm-agnostic} way and give an example. Based on the example, we derive a basic rule for the in-place versioning.

Before the main computation loop, 
the dual version allocates an extra copy of the target data objects (a new version). Then, in each iteration of the main computation loop, both versions of the data objects are involved into the computation, but memory write operations only happen to one version of the data objects (which we call ``working version''),
the other version (which we call ``consistent version'')
 remains unchanged until the next iteration.
At the end of each iteration, the working version 
is flushed out of the cache and becomes consistent in NVM. 
This version will not be changed in the next iteration, and 
becomes the consistent version since then.
The previous consistent version becomes the working version,
and is updated by memory write operations of the application.
Two versions alternate roles across iterations, with one version being consistent and the other being updated.
Hence, we ensure that there is always a consistent version in NVM for restart.
The recomputation is limited to at most one iteration, equivalent
to the recomputation in the frequent checkpoint we discuss in Section~\ref{sec:prelim_design}.

Figure 8 shows an example 
to further explain the basic idea. In this example, the array \textit{u} is the target data object. 
In the main loop (Lines 13-17) of the original code, all elements of $u$ are updated, and those elements are both read and written in each iteration of the main loop. 
In the dual version, 
we allocate an extra copy of $u$ ($u\_e$) and rename the original copy as $u\_o$. 
$u\_o$ is enforced to be consistent in NVM (Lines 4-6) before the computation loop. 
In the main loop, both $u_o$ and $u_e$ participate in
the computation. 
However, at any iteration, only one version of $u$
is updated, and the other version is read.
The update to one version of $u$ is naturally embedded
in the place of write operations (Line 12).
Also, at any iteration, we always maintain a consistent version of $u$ in NVM.
Depending on the iteration number (odd or even), we decide
which version should be updated and which one should be consistent.
The two versions switch their roles (either write or read) after each iteration (Lines 19-25).

Based on the above example and description in an algorithm-agnostic
way, we derive a basic rule for our in-place versioning.

\begin{itemize}
\vspace{-10pt}
\item Basic rule: within each iteration of the main computation loop, write operations happen to one version of the target data objects and read operations happen on the other version. Alternate the role of the two versions, and flush data blocks of the updated version out of caches after each iteration.  
\end{itemize}

\definecolor{codegreen}{rgb}{0,0.6,0}
\definecolor{codegray}{rgb}{0.5,0.5,0.5}
\definecolor{codepurple}{rgb}{0.58,0,0.82}
\definecolor{backcolour}{rgb}{0.95,0.95,0.92}

\lstdefinestyle{style1}{
    commentstyle=\color{codegreen},
    keywordstyle=\color{magenta},
    numberstyle=\tiny\color{codegray},
    stringstyle=\color{codepurple},
    basicstyle=\footnotesize,
    numbers=left,                    
    numbersep=5pt, 
	escapeinside={(*@}{@*)},
}
\lstset{style=style1}

\begin{table}
\centering
\small
\begin{tabular}{c p{0.1cm} c}
  \begin{lstlisting}[language=c++]
...
//initialization of u[]
init(u);

void update(u) {
  ... 
  for(i=0;i<Nu;i++) 
    u[i] = u[i] + e;
  ...
}

//main computation Loop  
for (it=0; it<Nit;i++){  
  ...
  update(u);
  ...
}
...
  (a) The original code
\end{lstlisting}
& 
& \begin{lstlisting}[language=c++]
...
/*u_o and u_e are 
two versions of u*/
u_e = malloc(..);
init(u_o);  
flush_cache(u_o); 

void update(u_new, u_old) 
{
  ... 
  for(i=0;i<Nu;i++) 
    u_new[i] = u_old[i] + e;
  ...
}

//main computation loop
for (it=0; it<Nit;i++){
  ...
  if (it%2==0) {
    update(u_e,u_o);
    flush_cache(u_e);
  } else {
    update(u_o,u_e);
    flush_cache(u_o);
   }
  ...
}     
...
(b) The dual version
\end{lstlisting}\\
\end{tabular}
\caption*{Figure 8: An example to explain the basic idea of the dual version described in an algorithm-agnostic way. $u$ (an array) is the target data object. $u$ has $Nu$ number of elements.}
\label{table:basic_rule}
\vspace{-20pt}
\end{table}

Although the basic rule is straightforward, it can be applied to many 
target data objects (see Table 2). However, the basic rule is also
very restricted. 
There are two special cases violating the basic rule.
In the first case, within one iteration, read operations reference one version (i.e., the consistent version) before any update happens to the target data object.
However, after the first update, read operations should reference the updated version (i.e., the working version) for program correctness. Read operations should not use the same version before and after the first update.
We name this case as \textit{post-update version switch for read operations}. We use an example to further explain it. 

\textbf{Special case I: post-update version switch for read operations.}
See Figure 9. In this example, we only show the routine where the updates to the target data object (the array $u$) happen (the routine \textit{update}), but ignore the main computation loop which is already shown in Figure 8.

In this example, 
for the first update of $u$ (Line 4 in Figure 9.b), we can 
use the basic rule correctly. The read operations use $u\_old$.
However, after the first update (Line 6 in Figure 9.b), we should read the most recent update from $u\_new$, not $u\_old$ suggested by the basic rule (see Line 6 in Figure 9.c for a correct version).
The read operations in Lines 4 and 6 in Figure 9.c use different version of $u$ after the first update in Line 4 in Figure 9.c. 

%
\lstset{style=style1}

\begin{table*}
\centering
\begin{tabular}{c p{1cm} c p{1cm} c}
   \begin{lstlisting}[language=c++]
void update(u) {
  ...
  for(i=0;i<Nu;i++) {
    u[i] = u[i] + e;
    ...
    u[i] = u[i] + f;
    ...
  }
  ...
}     

(a) The original code
\end{lstlisting}
&
& \begin{lstlisting}[language=c++,escapechar=!]
void update(u_new, u_old) {
  ... 
  for(i=0;i<Nu;i++) {
    u_new[i] = u_old[i] + e;
    ...
    u_new[i] = u_old[i] + f;
    ...
  }
  ...
}
(b) The wrong code based on 
the basic rule
\end{lstlisting} 
&
& \begin{lstlisting}[language=c++]
void update(u_new, u_old) {
  ... 
  for(i=0;i<Nu;i++) {
    u_new[i] = u_old[i] + e;
    ...
    u_new[i] = u_new[i] + f;
    ...
  }
  ...
}      

(c) The correct code 
\end{lstlisting}\\
\end{tabular}
\caption*{Figure 9: Special case I: post-update version switch for read operations. The target data object is $u$. The main computation loop is ignored in this figure. $u$ has $Nu$ number of elements. Line 6 in Figure 9.b is the incorrect code.}
\label{table:special_case_I}
\vspace{-20pt}
\end{table*}

The other case violating the basic rule is that elements of the target data object
are not updated uniformly within an iteration.
As a result, read operations should reference one version for
some elements of the target data object, but reference the other
version for the other elements. 
We use an example to further explain it.

\textbf{Special case II: nonuniform updates.} 
Figure 10 gives an example. 
There are two loops in the figure, each of which updates $u$.
In the first loop (Figure 10.a), the elements from 1 to $Nu-2$ of $u$ are updated,
while the elements 0 and $Nu-1$ are not updated.
In the second loop, all elements are updated.
Hence, across two loops, all elements are not updated uniformly.

Based on the basic rule, we replace $u$
in the first loop with the two versions of $u$ (Line 4 in Figure 10.b), which is correct.
In the second loop, 
we do the same thing (Line 7 in Figure 10.b) based on the basic rule. 
However, the program will not run correctly.
For the elements $u[0]$ and $u[Nu-1]$ that have not been updated in the first loop, we should use $u\_old$ for read operations in the second loop (Line 8 in Figure 10.c), while for the other elements that have been updated, we should use $u\_new$ for read operations (Line 10 in Figure 10.c). 

\lstset{style=style1}
\begin{table*}
\centering
\begin{tabular}{c p{1cm} c p{1cm} c}
   \begin{lstlisting}[language=c++]
void update(u) {
  ...
  //The first collective 
  //update to u
  for(i=1;i<Nu-1;i++) 
    u[i] = u[i] + e;
  ...
  //The first collective 
  //update to u
  for(i=0;i<Nu;i++) 
    u[i] = u[i] + f;
  ...
}
(a) The original code

  \end{lstlisting}
&
& \begin{lstlisting}[language=c++]
void update(u_new, u_old) {
  ...
  for(i=1;i<Nu-1;i++) 
    u_new[i] = u_old[i] + e;
  ...
  for(i=0;i<Nu;i++) 
    u_new[i] = u_old[i] + f;
  ...
}    



(b) The wrong code based on 
the basic rule
\end{lstlisting} 
&
& \begin{lstlisting}[language=c++]
void update(u_new, u_old) {
  ...
  for(i=1;i<Nu-1;i++) 
    u_new[i] = u_old[i] + e;
  ...
  for(i=0;i<Nu;i++) {
    if (i==0 || i==Nu-1)
      u_new[i] = u_old[i] + f;
    else
      u_new[i] = u_new[i] + f;
  }
  ...
}      
(c) The correct code
\end{lstlisting}\\
\end{tabular}
\caption*{Figure 10: Special case II: the elements of the data object $u$ are not updated uniformly. The main computation loop is ignored in this figure. $u$ has $Nu$ number of elements. Line 7 in Figure 10.b is the incorrect code.}
\label{table:special_case_II}
\vspace{-20pt}
\end{table*}

To handle the above two cases and enable automatic code transformation to implement the in-place versioning, we introduce a profile-guided code transformation.
This method uses the results of a profiling test to detect the first update and nonuniform updates, and then transforms the application into the in-place versioning accordingly.
We particularly target on arrays, the most common target data object in HPC applications. 
We explain our method in details as follows.

Our method first leverages an LLVM compiler~\cite{Lattner:Mthesis} instrumentation pass~\cite{ispass13:shao} to generate a set of dynamic LLVM instruction traces for the first iteration of the main computation loop. 
Those traces include dynamic register values and memory addresses referenced in each instruction. 
Each of the traces corresponds to either a loop or instructions
between two neighbor loops.
For example, the \textit{update} routine in Figure 10.a has three traces:
Two of them correspond to {\fontfamily{qcr}\selectfont for} loops and the third one corresponds to the instructions between the two loops.
We also record the whole memory address ranges of the target data objects in the beginning of each trace, based on the LLVM instrumentation.

Furthermore, we develop a trace analysis tool. Given the traces and memory address ranges of the target data objects as input, this tool tracks register allocation and memory references to determine which elements are updated in each trace.
Based on the analysis results across and within the traces, we identify the first update for each target data object; 
we also determine the coverage of each loop-based update (e.g., Lines 7-8 in Figure 8.a) and whether the coverages in all loop-based updates are different. This will be used to detect non-uniform update.

Based on the trace analysis results, we use a static LLVM pass to 
replace the references to the target data objects with 
the references to either the working version or the consistent version.
In particular, any read reference to the target data object before the first update will be replaced with the reference to the old version of the target data object (i.e., the consistent version); after the first update,
any read reference to the target data object will be replaced with
the reference to the new version (i.e., the working version). Any write reference to the target data object is always replaced with the reference to the new version, based on the basic rule.
Figure 9.c is an example of such replacement. 

If nonuniform updates are detected, then for a loop-based structure we need to add control flow constructs within the loop to control which version of the data objects should be used.
Figure 10.c (Lines 7-10) is such an example. 
However, in practice, we find that such control flow constructs can be rather sophisticated, especially for a statement of the loop with multiple elements of the data objects. Furthermore, 
the prevalence of such control flow constructs in loops can bring
large performance overhead. 
Hence, we do not apply the in-place versioning to the data object with nonuniform updates.
Instead, we use our preliminary design 2 (i.e., data copying based on non-temporal load/store) at the persistence establishment point for those target data objects.

\textbf{Discussion.}
We profile the first iteration to detect the first update and nonuniform updates.
This method aims to generate a short trace and make the trace analysis time manageable.
This method is based on an assumption that the first iteration and the rest of
iterations in the main loop have the same read and write patterns for the target data objects.
Based on our experience with 10 data objects from six NPB benchmarks (24 input problem sizes) and 7 data objects from a large-scale production code (Nek5000), we find such assumption is true in all cases.

Furthermore, we find that different input problems (not different input problem size) can have different read and write patterns to the data objects, and hence needs to generate different code for the in-place versioning.
However, profiling the first iteration and generating the code is quick, based on our compiler-based approach.

\textbf{In-place versioning vs. checkpoint.} 
There is a significant difference between the in-place versioning and checkpoint mechanism. 
Creating data copy in the checkpoint mechanism is an extra operation,
and also the data copy is not involved in the computation; 
Creating data copy in the in-place versioning 
leverages inherent memory write operations in the application,
and is part of the computation (not extra operation).
Hence, the in-place versioning significantly reduces data copying overhead
from which the checkpoint mechanism suffers. 

However, the in-place versioning can bring performance loss from two perspectives. First, the in-place versioning has to allocate one extra data copy before the main computation loop. However, this cost happens just once, and can be easily amortized by the main computation.
Second, the in-place versioning increases memory footprint of the application, because the two versions of the target data object are involved in the computation. This may increase CPU cache miss rate, which hurt performance. This may also consume more DRAM cache space,
reducing the DRAM space for other data objects.
However, we see small performance difference (less than 8.2\% and 2.7\% on average) between the in-place version and the native execution without it. The reason is as follows.

For the DRAM cache problem, the software-based cache management we use 
in our study~\cite{eurosys16:dulloor} treats each extra data copy as a new data object and chooses the best data placement in DRAM and NVM for optimal performance, which effectively reduces the impact of larger working set in the in-place versioning.  
For the CPU cache problem, we study it based on performance counters, but do not find significant increase in cache miss rates because of the ``streaming-like'' memory access patterns in target data objects. We discuss it further in the performance evaluation section.

\subsection{Optimization of Cache Flushing}
\label{sec:perf_opt}
The in-place versioning avoids memory copying. However, to make data consistent between NVM and caches at the persistence establishment point, we need to flush caches.
As shown in Figure~\ref{fig:perf_loss_breakdown}, periodically flushing caches accounts for a large portion of the total overhead. 
The fundamental reason for such large overhead is that we cannot know which cache blocks of the data objects are in the cache hierarchy and whether they are dirty, and have to issue cache flushing instructions on every single cache block of the target data objects. 

To reduce the cache flushing cost, we propose two optimization techniques: whole cache flushing and proactive cache flushing.

\textbf{Whole cache flushing.}
The basic idea of the whole cache flushing is to use {\fontfamily{qcr}\selectfont WBINVD} instruction to flush the entire cache hierarchy, instead of flushing individual cache blocks of the target data objects. If the size of the target data objects is much larger than the last level cache size, 
it is highly possible that most of the cache blocks are not in caches, and flushing the entire cache hierarchy is cheaper than flushing all cache blocks of the target data objects. 

However, {\fontfamily{qcr}\selectfont WBINVD} is a privileged instruction, and only the kernel level code can issue this instruction. Hence we introduce a kernel module 
that allows the application to indirectly issue the instruction. The drawback of using {\fontfamily{qcr}\selectfont WBINVD} is that the cache blocks that do not belong to the target data objects
are flushed out of the caches. If those cache blocks are to be reused, they have to be
reloaded, which lose performance. However, when the total size of the target data objects is large enough,
flushing all cache blocks of the target data objects that are not resident in caches is much more expensive than data reloading because of {\fontfamily{qcr}\selectfont WBINVD}. We empirically decide that
if the total size of the target data objects is ten times larger than the last level cache size, it is beneficial to use {\fontfamily{qcr}\selectfont WBINVD}.

\textbf{Asynchronous and proactive cache flushing.}
In the in-place versioning, we trigger cache flushing (including CPU cache flushing with {\fontfamily{qcr}\selectfont WBINVD} and DRAM cache flushing)
at the persistence establishment point to make the working version consistent in NVM.
To improve cache flushing performance, we want to remove cache flushing off the execution critical path as much as possible. 
Also, we can trigger cache flushing ahead of the persistence establishment point
under certain conditions (discussed as below). 
We introduce a helper thread-based mechanism to implement asynchronous
and proactive cache flushing. 

In particular, we do not wait until the persistence establishment point to flush caches. Instead, as soon as the working version is not updated in the current iteration, a helper thread will
proactively flush caches. Furthermore, the cache flushing does not have to be finished at the end of each iteration. As long as the working version from the last iteration 
is not read in the current iteration, the cache flushing can continue. But, the helper thread must finish cache flushing
at the point where the working version 
from the last iteration is read for the first time.
Figure 11 describes the idea.


\begin{figure}
\centering
\includegraphics[width=0.5\textwidth, height=0.3\textheight]{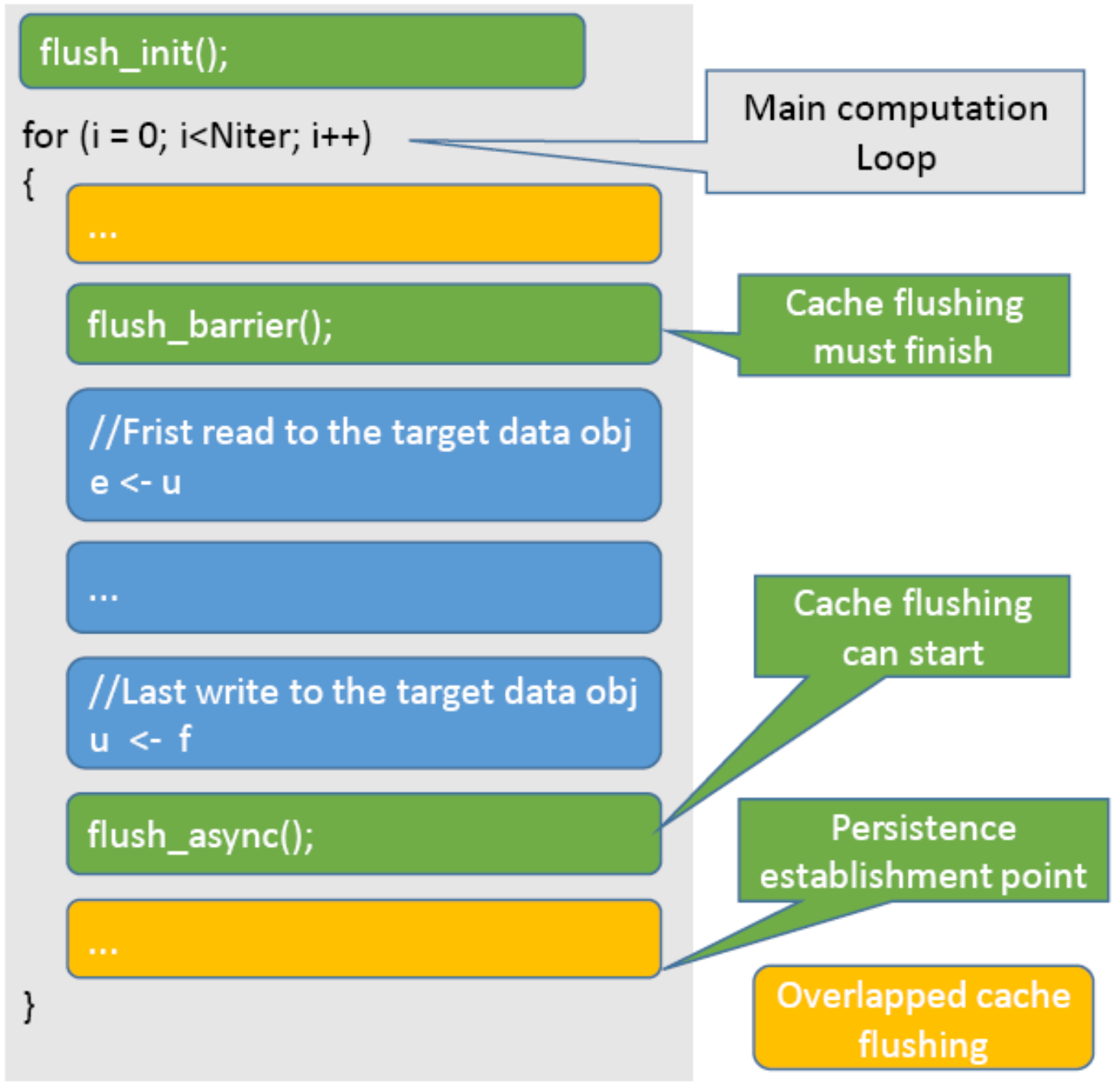}
\vspace{-20pt}
\caption*{Figure 11: The proactive cache flushing scheme.}
\label{fig:proactive_cache_flushing}
\vspace{-20pt}
\end{figure}

To implement the above proactive cache flushing, we develop a lightweight library for HPC applications and a set of APIs. To use the library, the programmer needs to insert a thread creation API (flush\_init() in Figure 11) before the main loop to create a helper thread and a FIFO
queue shared between the helper thread and main thread.
The programmer also needs to insert an API (flush\_async() in Figure 11) into the program to specify where the cache flush can happen within each iteration; 
The cache flush point does not have to be the same as the persistence establishment point. 
Using this API will insert a cache flush request into the FIFO queue.
The programmer also needs to insert an API to specify where the cache flush must finish within each iteration (flush\_barrier() in Figure 11). This API works as a synchronization between the helper thread and the main thread to ensure that the working version is completely flushed before it becomes the consistent version and read by the application.

\textbf{Discussion.} Similar to any help thread-based approaches~\cite{mem_reg_cf11, ipdps10:tiwari, hpca03:mutlu, tpds09:prefetching}, our approach depends on the availability of 
idling core for helper threads. We expect that the future many-core platform can provide such core abundance. Note that even without the helper thread, the in-place versioning with {\fontfamily{qcr}\selectfont WBINVD} already provide significant performance improvement over checkpoint, shown in Figure 12 in the evaluation section.


\section{Evaluation}
\label{sec:eval}

We evaluate the in-place versioning (IPV) in this section.
Unless indicated otherwise, IPV includes optimized cache flushing 
and helper thread in this section.
Also, the data persistence establishment happens at every iteration of the main computation loop, which aims to build
high resilience and minimize recomputation for future HPC.
We use the native execution, which has neither checkpoint
nor IPV, as our baseline. 
An ideal performance of IPV should be close to that of
the native execution as much as possible.

We study the performance on two test platforms. 
One test platform is a local cluster. Each node of it has two eight-core Intel Xeon E5-2630 processors (2.4 GHz) and 32GB DDR4. We use this platform for tests in all figures except Figure~\ref{fig:prelim_design1} in Section~\ref{sec:prelim_design}. 
We deploy Quartz on such platform to emulate a heterogeneous NVM/DRAM system with NVM configured with 1/8 DRAM bandwidth and DRAM configured with 256MB capacity to enable a practical emulation of NVM~\cite{eurosys16:dulloor, NVMDB}. 
The other test platform is the Edison supercomputer at Lawrence Berkeley National Lab (LBNL). We use this platform for tests in Figure~\ref{fig:prelim_design1}. 
Each Edison node has two 12-core Intel Ivy Bridge processor (2.4 GHz) with 64GB DDR3. 
We cannot install Quartz on Edison to enable a practical emulation of NVM, because Quartz requires a privileged access to the system. Hence, we perform most of the tests on the local cluster.

We use six NPB benchmarks (CLASS C) and one production application (Nek5000) with the eddy input problem ($256 \times 256$). 
Table 2 gives more information on the benchmarks and application. The table also lists how the target data objects are transformed into IPV based on either basic rule, post-update version switch, or nonuniform update.
For NPB benchmarks, the target data objects are chosen based on typical checkpoint cases, algorithm knowledge, and benchmark information.
For Nek5000, the target data objects are determined by the checkpoint mechanism in Nek5000. 

\begin{table}
\centering
\footnotesize 
\caption*{Table 2: Target Data objects for checkpointing and the in-place versioning. IPV in the table stands for the in-place versioning.}
\vspace{-10pt}
\begin{tabular}{|p{1.0cm}|p{1.1cm}|p{1.3cm}|p{1.9cm}|p{1.5cm}|}
       \hline
       \textbf{Bench-mark} & \textbf{Data obj} & \textbf{IPV (basic rule)} & \textbf{IPV (post-update version switch)} & \textbf{IPV (nonuniform update)} \\
        \hline \hline
      FT & u0,u1,u2 & u1,u2 & u0 & -  \\  \hline
      CG & p,r,z & p & r,z & -    \\  \hline
        BT & u & u & - & -    \\  \hline
        SP & u & u & - & -    \\  \hline
        LU & u & u & - & -    \\ \hline
        MG & r & - &  -& r   \\ \hline
        Nek5000 (eddy) & vx, vy, vz, pr,xm1,ym1, zm1 & pr,xm1,ym1, zm1 & vx,vy,vz & -  \\ \hline
\end{tabular}
\label{tab:benchmarks}
\vspace{-20pt}
\end{table}

Figure 12 compares the performance of 
the baseline, the preliminary design 2 (i.e., checkpoint with cache bypassing), IPV with neither cache flushing nor helper thread, IPV with cache flushing (no helper thread), and IPV with everything.
Comparing with the baseline, IPV achieves rather small runtime overhead (4.4\% on average and no larger than 9.5\%). Most of the performance improvement comes from the removal of data copying. In particular, regarding IPV (no cache flushing and helper thread) and the preliminary design 2, both of them do not have cache flushing, but IPV (no cache flushing and helper thread) performs 9\% better on average because of no data copying. This fact is especially pronounced in Nek5000, where IPV (no cache flushing and helper thread) performs 26\% better than the preliminary design 2. 

Furthermore, IPV cannot be applied to
MG because of nonuniform updates (see Table 2).
Hence MG does not have performance data for any IPV.
However, MG with the helper thread to enable proactive and asynchronous data copying in the figure has 5.4\% performance improvement over the preliminary design 2.

\begin{figure*}
\centering
\includegraphics[width=1.0\textwidth, height=0.2\textheight]{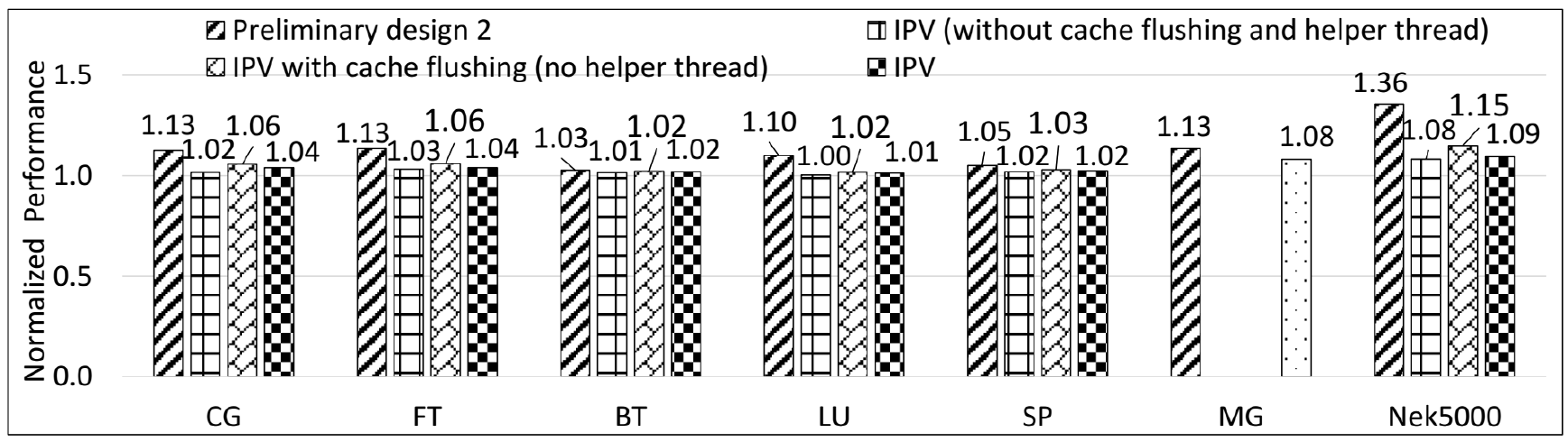}
\vspace{-20pt}
\caption*{Figure 12: Performance difference between the native execution (baseline), the preliminary design 2 (checkpoint with cache bypassing), and different IPV cases. Performance is normalized to that of the native execution. MG does not have the results for IPV. The dotted bar in MG is the case of checkpoint with a helper thread for asynchronous and proactive data copying.}
\label{fig:ipv_perf}
\vspace{-10pt}
\end{figure*}

To further study the performance of IPV, we focus on the performance difference between IPV without cache flushing and IPV. We aim to study the effectiveness of proactive and asynchronous cache flushing. 
In Figure 13, we measure performance of {\fontfamily{qcr}\selectfont WBINVD} and DRAM cache flushing, and quantify their contribution
to the total overhead (i.e., {\fontfamily{qcr}\selectfont WBINVD} plus DRAM cache flushing) in IPV.  The table below the figure quantifies how much of  the total overhead is overlapped with the application execution by the proactive and asynchronous cache flushing. 

Figure 13 reveals that the proactive and asynchronous cache flushing is pretty effective to hide the cache flushing overhead (or data copying for MG). At least 41\% of the total overhead is overlapped in all benchmarks. 
The non-overlapped cache flushing time is exposed to the application critical path and causes the performance difference between IPV and the native execution in Figure 12.

\begin{figure}
\centering
\includegraphics[width=0.5\textwidth, height=0.13\textheight]{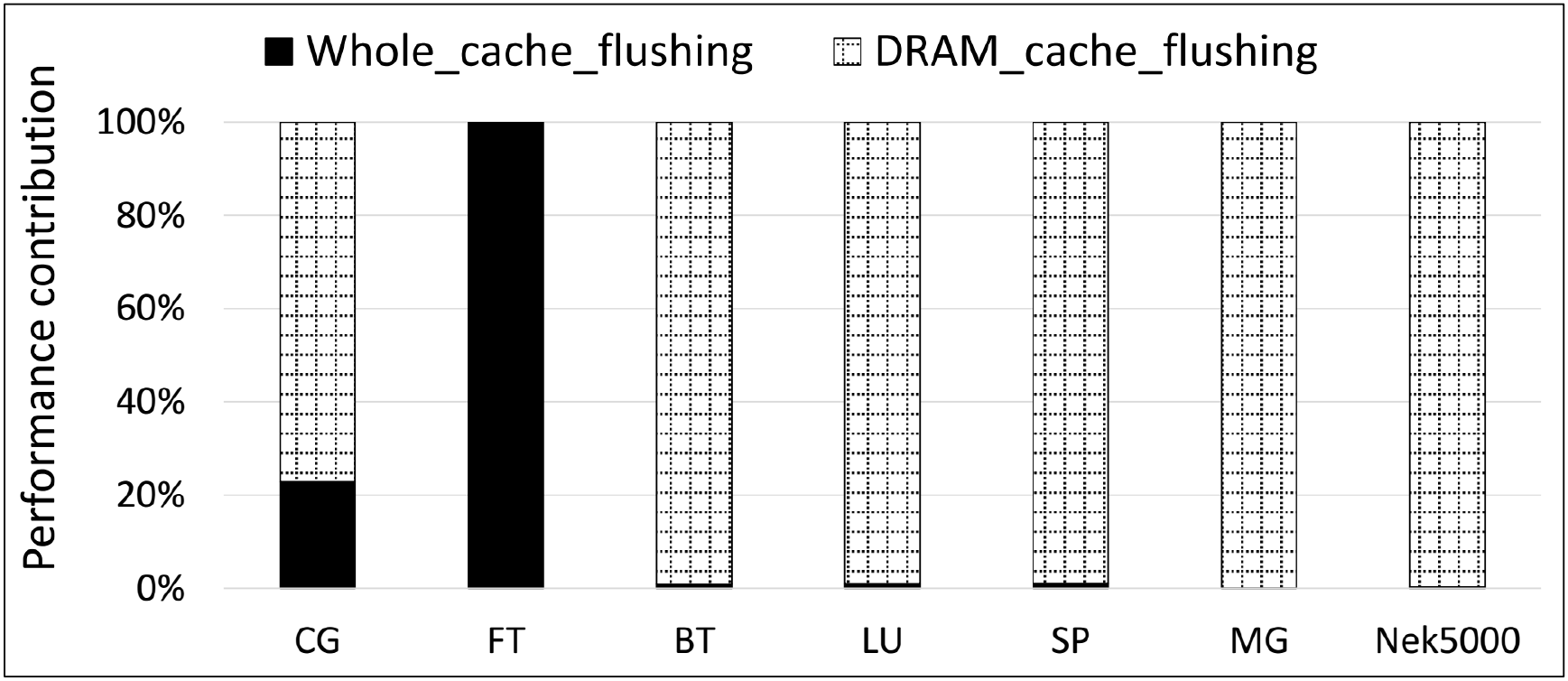}

\includegraphics[width=0.5\textwidth, height=0.03\textheight]{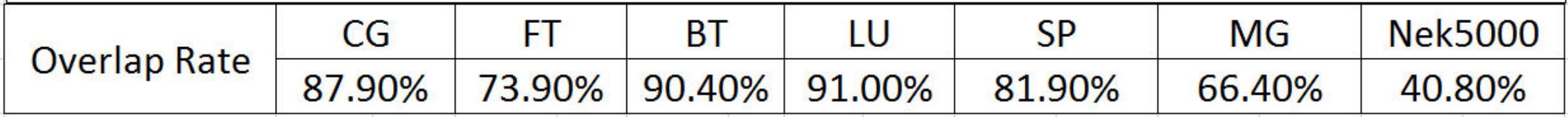}
\vspace{-20pt}
\caption*{Figure 13: Breakdown of the performance difference between the in-place versioning and in-place versioning without cache flushing.}
\label{fig:ipv_perf_breakdown}
\vspace{-10pt}
\end{figure}

IPV can cause extra CPU cache misses, because of two reasons. (1) The two versions of the target data objects increase working set size of the application; (2) {\fontfamily{qcr}\selectfont WBINVD} flushes the entire cache hierarchy. 

We measure the system-wide last level CPU cache miss rate for the native execution and IPV. Figure 14 shows the results. In general, we do not see big difference (up to 4\%) between the two cases in terms of the last level cache miss rate. This further explains the small performance loss between IPV and the native execution in Figure 12.

The reason that accounts for such small difference in the last level miss rate is as follows. {\fontfamily{qcr}\selectfont WBINVD} happens only once in each iteration, hence its impact on cache misses is not frequent. The two versions do increase the working set size of the application. However, within the original application, the target data objects are typically updated in a loop (e.g., the loop structure in \textit{update} routine in Figures 9 and 10) and there is
little data reuse across iterations of the loop. Such updates tend to be ``streaming-like'', which is not sensitive to the increase of working set size.
\vspace{-10pt}

\begin{figure}
\centering
\includegraphics[width=0.5\textwidth, height=0.13\textheight]{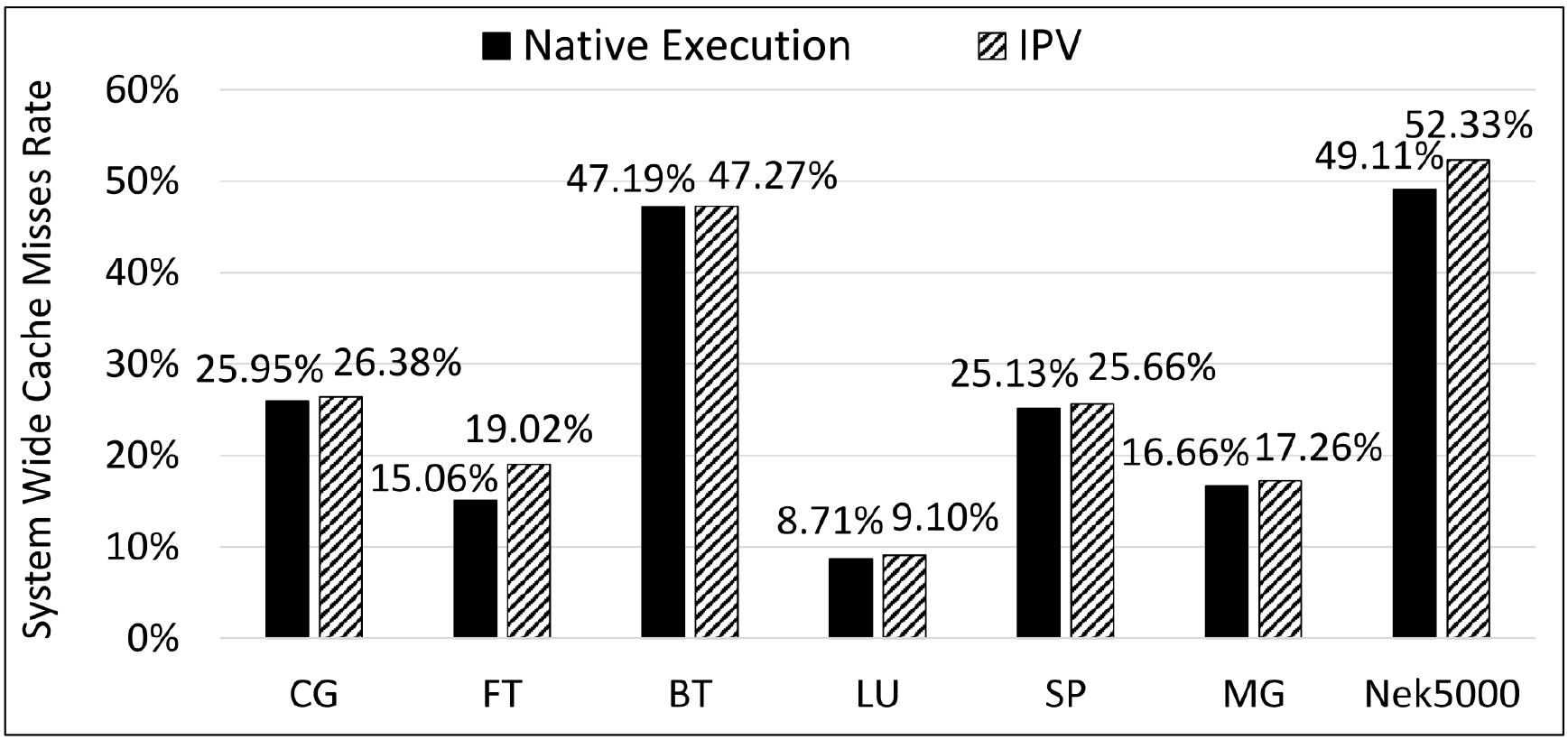}
\vspace{-20pt}
\caption*{Figure 14: Last level CPU cache miss rate difference between the baseline and the in-place versioning (no cache flushing).}
\label{fig:cache_miss_rate}
\vspace{-10pt}
\end{figure}

\section{Related Work}
\label{sec:related_work}
\begin{spacing}{0.9}
\textbf{Persistent memory.}
NVM has been explored to implement checkpoint as main memory.
Kannan et al.~\cite{ipdps13:kannan} use NVM only for checkpoint (not computation). To improve performance, they proactively move checkpoint data from DRAM to NVM before checkpoint is started. 
Gao et al.~\cite{ics15:gao} use a hardware-based approach to utilize runtime idling to write checkpoint and spread it across memory banks for load balance.
Ren et al.~\cite{micro15:ren} dynamically determine checkpoint granularity (cache block level or page level) based on memory update density.
Dong et al.~\cite{sc09:dong} introduce 3D stacked NVM and incremental
checkpoint to reduce checkpoint overhead.
Those prior efforts focus on good performance of NVM to establish persistence (checkpoint) in NVM, while we focus on how to maximize the benefit of non-volatility of NVM. Different from those prior efforts, our work avoids data copying, 
and does not require hardware assist.

To enable data consistence in NVM, many research efforts explore how to enforce write-ordering with minimum overhead. The epoch-based approach~\cite{sosp09:condit, Pelley:isca14, micro15:joshi, micro16:kolli} is one of those research efforts. This approach divides program execution into epochs, within which stores are allowed to happen concurrently without disturbing data consistence in NVM.
In fact, our proactive cache flushing (Section~\ref{sec:perf_opt}) is one variation of epoch. 
From the point where the cache flush happens to the point where the working version becomes the consistent version is an epoch where concurrent, persistent writes can happen.
However, most of the existing work is hardware-based and requires hardware support to implicitly identify epochs. Also, to apply the existing work to establish data persistence in HPC still needs a mechanism to maintain two versions of the target data objects. Our work requires no hardware support and the in-place versioning provides the two versions.

Some work explores redo-log and undo-log based approaches to build transaction semantics for data consistence in NVM.
This includes hardware logging~\cite{pm_iccd14, hpca17:joshi, stable_tr16:zhao}.
However, those approaches come with extensive architecture modifications.

There are also software-based approaches that introduce certain program constructs to enable data persistence in NVM~\cite{mnemosyne_asplos11, intel_nvm_lib, usenix13:rudoff, nv-heaps_asplos11, vldb_endow15:chatzistergiou, hpdc16:denny}. 
To use those program constructs, one have to make changes to OS and applications. 
The application can suffer from large overhead because of frequent runtime checking
or data logging. Our experiences with~\cite{intel_nvm_lib} show that CG and dense matrix multiplication 
suffer from \textbf{52\%} and \textbf{103\%} performance loss because of frequent data logging operations.
Our work in this paper has very small runtime overhead and does not require changes to OS.

\textbf{Checkpoint mechanism.}
Diskless checkpoint is a technique that uses DRAM-based main memory and available processors
to encode and store the encoded checkpoint data~\cite{tpds98:plank, Lu:2005:SDC:1145057, ppopp17:tang, isftc94:plank}. 
Because of the DRAM usage and the limitation of encoding techniques, diskless checkpoint has to leverage multiple nodes to create redundancy and only tolerates  up to a certain number of node failures.
Our method is a diskless-based approach, but leveraging non-volatility of NVM.
Our method does not have node-level redundancy in diskless checkpoint, and is independent of the number of node failures.

Incremental checkpoint is a method that only checkpoints modified data to save checkpoint size
and improve checkpoint performance~\cite{isftc94:plank, ics04:agarwal, icpads10:wang, ipdps09:bronevetsky}.
However, for those applications with intensive modifications between
checkpoints (e.g., HPL~\cite{ppopp17:tang}), the effectiveness of the incremental checkpoint method can be limited. 

Multi-level checkpoint is a method that saves checkpoint to fast devices (e.g., PCM and local SSD) 
in a short interval and to slower devices in a long interval~\cite{sc10:moody, sc09:dong, sc11:gomez}. 
By leveraging good performance of fast devices, the multi-level checkpoint removes expensive memory copy on slower devices. 
However, it can still suffer from large data copy overhead on fast devices, when
the checkpoint data size is large.
Our work introduces the in-place versioning to 
remove data copy by leveraging application-inherent write operations to update checkpoint data.
Hence, our method does not have the limitation of incremental and multi-level checkpoints.
\vspace{-10pt}
\end{spacing}



\section{Conclusions}
\begin{spacing}{0.9}
With the emergence of NVM, how to leverage performance and non-volatility characteristics of NVM for future HPC is largely unknown.
In this paper, we study how to use NVM to build data persistence for critical data objects of applications to replace traditional checkpoint.
Our study enables the frequent establishment of data persistence on NVM
with small overhead, which enable high resilient HPC and minimized recomputation.
\end{spacing}   

\bibliographystyle{ACM-Reference-Format}
\bibliography{li}  

\end{document}